# The electronic structure formation of $Cu_xTiSe_2$ in a wide range (0.04 < $x$ < 0.8) of copper concentration.

Shkvarin A.S.[1], Yarmoshenko Yu.M.[1], Yablonskikh M.V.[2], Merentsov A.I.[3], Shkvarina E.G.[1], Titov A.A.[1,3], Titov A.N.[1,3]

1 Institute of Metal Physics, Russian Academy of Sciences-Ural Division, 620990 Yekaterinburg, Russia

2 Sincrotrone Trieste S.C.p.A., Area Science Park S.S 14 Km 163.5, 34012 Basovizza, Italy.

3 Ural Federal University, 620089, Yekaterinburg, Russia

An experimental study of the electronic structure of copper intercalated titanium dichalcogenides in a wide range of copper concentrations ($x$ = 0.05 - 0.6) using the x-rays photoemission spectroscopy, resonant photoemission and x-rays absorption spectroscopy has been performed. Negative energy shifts of the Ti 2p and Se 3d core levels spectra and a corresponding decrease of the photon energy in Ti 2p absorption spectra with increasing concentration of copper have been found. Such sign-anomalous shifts may be explained by the shielding effect of the corresponding atomic shells as a result of the dynamic charge transfer during the formation of a covalent chemical bond between the copper atoms and the $TiSe_2$ matrix.

## 1. Introduction

The compounds $Cu_xTiSe_2$ were extended studied previously only for low concentrations of the copper atoms. Superconductivity as well as charge-density-wave state were discovered in these compounds at low copper concentrations [1]. In the first studies it was reported that the solubility limit of copper in the titanium diselenide is 11 at.% [1]. It was found later that the solubility limit is much higher and reaches 75 at. %, and the lattice parameter $c$ increases with increasing $x$ up to $x$ = 0.5 and starts to decrease after that with continuing increase of copper content [2]. Such a behavior is not typical for titanium dichalcogenides intercalated with the noble and 3d transition metals and was still observed only in $1T-Mn_xTiSe_2$ [3].

In general, three regions in the concentration dependence of the lattice parameter $c$ within the solubility region of copper in $TiSe_2$ can be distinguished: 0 < $x$ < 0.33, where a linear increase of the parameter $c$ with increasing $x$ is observed; 0.33 < $x$ < 0.5, where the growth of $c(x)$ turns to the saturation regime; $x$ > 0.5, where the increase of $x$ reduces $c$ [2]. The nature of the chemical bond between copper and $TiSe_2$ lattice has been previously studied only for the region $x$ < 0.33. It is obvious that the complex behavior of the $c(x)$ function reflects changing of the nature of the chemical bond between copper and $TiSe_2$ lattice. The present work is devoted to the study of this problem.

As it was established in [4] at low copper concentrations ($x < 0.33$) the shape of the Cu L absorption spectra corresponds to the free copper ions, i.e. copper atoms have $3d^{10}4s^2$ electronic configuration. The aim of this work is study of the $Cu_xTiSe_2$ electronic structure in a range of copper concentrations leading both to the lattice expansion and to the lattice contraction. For this purpose a wide range of the experimental techniques including x-rays photoemission spectroscopy (XPS), x-rays absorption spectroscopy (XAS) and resonant photoemission (ResPES) has been used. For the first time for the titanium diselenide intercalated with the 3d transition metal atoms (copper in the current work) a detailed joint analysis of the valence bands and the crystal structure in a wide concentration range of copper atoms has been carried out experimentally. These measurements were performed for samples with a small ($x = 0.05, 0.07, 0.09$), middle ($x = 0.33$) and large ($x = 0.58, 0.6, 0.8$) copper concentration.

## 2. Experimental techniques

Single crystals of $Cu_xTiSe_2$ ($x = 0.05, 0.07, 0.09, 0.33, 0.58, 0.6, 0.8$) were grown using direct sublimation technique in evacuated quartz ampoules from the polycrystalline phase of a given composition. The crystals are thin plates with dimensions about $2 \times 2 \times 0.1$ mm. The chemical composition was determined using the electron fluorescence analysis technique on the JEOL-733 spectrometer. The crystal structure was studied on the polycrystalline samples using x-rays diffraction technique in the Institute of Metallurgy UrD RAS, KUC "Ural-M" using diffractometer Shimadzu XRD 7000 C Maxima (Cu $K_\alpha$ radiation, graphite monochromator). All the materials have the same allotropic modification 1T with the space group P-3m1. The full-profile analysis was performed using the software package General Structure Analysis System (GSAS) [5]. The starting model was as follows: space group P-3m1, atomic coordinates Ti (0 0 0), Se (1/3 2/3 z), Cu (0 0 1/2), z ~ 0.254. The background was approximated by the Chebyshev polynomial with eight parameters, the thermal parameters were calculated in the anisotropic form, and the pseudo-Voight line profile function with five coefficients was used. In the calculation the unit cell parameters, z-coordinate of the selenium atoms and filling of the octahedral positions in the Van der Waals gap, in which copper atoms are located, were refined.

The measurements were carried out at the BACH [6] and CiPo [7] beamlines at the ELETTRA synchrotron and RGBL beamline at the BESSY synchrotron. To obtain the fresh surface all the samples were cleaved in a vacuum chamber at a pressure of $3-5 \times 10^{-9}$ Torr. Purity of the surface was confirmed by the absence of oxygen and carbon peaks at the survey spectra. The monochromator resolution was $E_{mono} = 0.12$ eV, the resolution of the photoelectron analyzer was set to $E_{pa} = 0.147$ eV, which gives a total resolution of $E_{tot} = 0.19$ eV. Calibration of the valence bands spectra was made by the gold spectrum in the vicinity of the Fermi energy. The absorption spectra $L_{2,3}$ (2p → 3d4s transition) of Ti and Cu were measured in the total-electron-yield mode and the energy was calibrated by the absorption spectrum of the pure metal. The spectra of the core levels were calibrated by Au $4f_{5/2}$ level ($E_b$ = 84 eV).

## Results and discussion

1. **Crystal structure**

Table 1 and Figure 1 give the lattice parameters and show the dependence of interatomic distances on the composition, correspondingly, of the five samples. Increase of the copper content leads to the decrease of the Se-Ti-Se package thickness and to the increase of the Se-Cu-Se package thickness. Herewith in the copper concentration range up to 33 at.% the thickness of the Se-Ti-Se package varies insignificant. Further increase of the copper concentration leads to a drastic decrease of the Se-Ti-Se package thickness. In the studied copper concentration range the relative change of the $c$ parameter is $\Delta c/\Delta a = 2$. Therefore, change of Ti-Se and Cu-Se bonds lengths occurs mainly in $c$ direction. At low copper concentration in the crystal structure of $Cu_xTiSe_2$ the structural fragments $TiSe_6$ and $TiSe_6$-Cu dominate (Figure 2). The increase of copper content leads to the increase of the concentration of $TiSe_6$-Cu clusters with one adjacent copper atom in their nearest environment. Further increase of the copper content increases the concentration of Cu-$TiSe_6$-Cu clusters. This process leads to the appearance of two crystallographically nonequivalent positions of Ti atoms in the Ti(1)$Se_6$-Cu and Cu-Ti(2)$Se_6$-Cu clusters coordinated with one and two copper atoms correspondingly (Figure 2). The concentration of clusters depending on the copper content is in good agreement with the data of magnetic measurements and with the lattice parameters values [2].

2. **Core levels photoemission spectra and absorption spectra.**

The Ti 2p photoemission spectra for all the samples were obtained at the same conditions; all the spectra have practically the same shape and are undistinguishable from the Ti 2p spectrum of the original $TiSe_2$ (Figure 3, left panel) [8]. However if copper content exceeds $x = 0.33$ the $\Delta E \approx 0.3$ eV reduction of the binding energy is observed. The shape of the Se 3d spectra of all samples is also similar to the shape of the Se 3d spectrum of pure $TiSe_2$ (Figure 3, right panel). The Se $3d_{5/2}$-$3d_{3/2}$ multiplet is well resolved. This circumstance indicates a good quality of the single crystals. For the Se 3d spectra as for the Ti 2p spectra a decrease of the binding energy that reaches $\Delta E \approx 1.3$ eV at $x = 0.6$ with the increase of the copper concentration is also observed. This unusual coincidence of the directions of the Ti 2p and Se 3d binding energy shifts requires special discussion. It should be noted that copper concentration $x = 0.33$ is the boundary between the region with a linear increase of the conduction electrons concentration with increasing $x$ ($x < 0.33$) and the region where the conduction electrons concentration is practically independent on $x$ ($x > 0.33$) [2]. This fact was interpreted in [9] as an evidence of the transition from the region with the dominance of van der Waals chemical bond to the region with the formation of the covalent chemical bond between copper and host lattice. Thus in the region with $x > 0.33$ the localization of the electrons introduced with intercalated copper near the Cu atoms is expected.

The shape of the Cu 2p spectrum for maximal copper concentration (Figure 4, inset of right panel) is similar to one for copper [10] (Cu in $3d^{10}4s^1$ state) and for copper oxide $Cu_2O$ [11] (Cu in $3d^{10}4s^0$ state). We can say with confidence that copper is here not in the divalent state.

The titanium absorption spectra for all copper concentrations are identical to the one in the pristine $TiSe_2$ (Figure 4, left panel) [8]. The unchanged shape of the titanium absorption spectrum even at high copper concentrations distinguishes these compounds from other titanium dichalcogenides intercalated with 3d metals: even small amount of intercalated 3d metal led to the observed change of the titanium absorption spectrum shape [12, 3]. However, when the titanium absorption spectrum is unchanged the gradual energy shift that reaches 0.3 eV at high copper concentration is observed (Figure 4, left panel). It is entirely consistent in magnitude and sign with a shift which is observed in the Ti 2p XPS.

The shape of Cu $L_{2,3}$ absorption spectra depends on the copper concentration (Figure 4, right panel). For small copper concentrations it was established that copper atoms behave itself as free atoms with a $3d^{10}4s^1$ configuration [4]. The increase of the copper concentration leads to the appearance of an additional peak B, shift of the peak C and increase of the intensity of the peak A on the copper absorption spectrum. The overall shape of the spectrum is at the same time unchanged. It gives the evidence of the manifestation of $3d^{10}4s^0$ configuration since the appearance of other configurations could be seen on the Cu 2p core level spectrum. The absorption spectra are in this case more informative than core level spectra.

During the study of core level spectra and absorption spectra the following experimental data were obtained:

1. A nonlinear decrease of the Ti 2p and Se 3d XPS binding energies with an increase of the copper concentration is observed keeping the shape of the XPS peaks unchanged;

2. A reduce of the absorption spectra Ti $L_{2,3}$ energy;

3. Cu 2p XPS stay unchanged; a change of the Cu $L_{2,3}$ absorption spectra shape is observed.

3. **The resonant XPS of the valence band.**

The valence band spectra in the Cu 2p-3d resonant mode both for small [4] and for high (Figure 5) copper concentrations are almost identical to the valence band spectra of the pristine TiSe$_2$. On the differential spectra the crossover that corresponds to the transition from the resonant Raman Auger peak (marked as RRAS on fig. 5) to the Auger-peak is clearly visible. It is easy to determine the energy at which the crossover is observed [13]. Change of the excitation energy leads only to the overall increase of the intensity of spectrum, as the constant initial state (CIS) spectrum shown in Figure 5 confirms. The absence of resonant behavior is caused by an insignificant spectral contribution of the copper atoms in this region of the excitation energy. The photoionization cross section of copper at $E_{ex}$ = 950 eV is two times less than that of selenium [14]. Bearing in mind concentration of elements in the compound, copper even at its higher concentration gives the contribution in the valence band on the order less than selenium.

The valence band spectra in Cu 3p-3d resonant excitation mode are on great interest since at the excitation energy $E_{ex}$ = 75 eV the photoionization cross section of copper is 66 times higher than that of selenium [14]. It's worth to note however, that the escape depth of photoelectrons on the Cu 3p absorption edge is several times less than that on Cu 2p edge and as is well known not exceeds 1-2 atomic monolayer. The valence band spectra of the compounds with high copper concentration ($x$ = 0.58) in the Cu 3p-3d resonant excitation mode (Figure 6, upper panel) differ both from the valence band spectrum of the pristine TiSe$_2$ [8] and from the valence band spectra of the compounds with low copper concentrations [4].

The terms of Cu$^0$ states are more clearly seen on the valence band spectra for compounds with high copper concentration than for compounds with low copper concentration [15, 16]. Analysis of the valence band spectra obtained in Cu 3p-3d resonant excitation mode shows that Cu 3d band is filled, since the shape of the spectrum keep unchanged with a change of the excitation energy near the absorption edge.

In contrast to Cu resonant mode the valence band spectra obtained in Ti 2p-3d resonant excitation mode show a strong dependence on the copper concentration (Figure 6, bottom panel). For low copper concentration the valence band spectra are similar to the spectra of the original TiSe$_2$ [17]. However even at the lowest copper concentration $x$ = 0.044 the resonant band with $E_B$ = 0.27 eV is observed which intensity increases together with the copper concentration $x$.

We believe that the electronic states described by this resonant band directly below the Fermi energy are s-type. The presence of the resonant contribution of Ti 3d states indicates the dynamic transfer of electrons between titanium and its environment. The transfer of electrons takes place with direct participation of selenium atoms due to hybridization of Cu 3d states with Se 4p states and Ti 3d states with Se 4p states as shown by model calculations of the electronic structure of other intercalated transition metal dichalcogenides: $Co_xTiSe_2$ [18] and $Cu_xNbSe_2$ [19]. It's important to note here that the interaction between titanium atoms and atoms of selenium and copper does not lead to additional filling of Ti 3d band since no changes of Ti $L_{2,3}$ absorption spectra is observed in the whole range of copper concentration. As mentioned above Ti $L_{2,3}$ absorption spectra are extremely sensitive to a stationary charge transfer [3, 8]. Therefore in terms of ionic chemical bonding an additional charge transfer on Ti does not occur. Negative shifts of the Ti2p and Se3d core levels binding energy can be explained by the dynamic charge transfer from the "small" systems (whose role is played by Cu-sublattice) to the "large" system (the $TiSe_2$ matrix). Here we use the concept of "small" and "large" systems proposed in [20] in describing the interaction of thin films and substrates with the surface of bulk materials. The charge transfer can be described as the tunneling process of redistribution of charge between "large" and "small" systems. This approach has been successfully used in [21] for the titanium diselenide intercalated with 3d metal atoms. The extent to which the dynamic charge transfer has an effect on the core levels and valence band photoelectrons can be estimated by comparing the lifetimes of the corresponding states-vacancies in the core levels and in the resonant bands located directly below the Fermi energy in the valence band spectra. The half-width of the corresponding lines is: 1.82 eV for Ti 2p, 0.42 eV for the resonant line R2 in the valence band. According to the Heisenberg uncertainty relation $\Delta\tau \times \Delta E \geq \hbar/2$ the lifetime (in this case of the excited state) $\Delta\tau$ is inversely proportional to the bandwidth. Obviously, in this case the lifetime of the charge transfer is much higher than the relaxation time of a core vacancy. It means that the core states are the bulk of time under the influence of the additional potential of electrons localized on the corresponding nuclear sites as a result of the dynamic charge transfer. This interpretation is also suitable for the absorption spectra. It can be concluded that the main factor determining the energy shifts of the absorption and core levels spectra is a shielding of the Ti and Se atomic shells by the electrons of the dynamical charge transfer.

### 4. Conclusion.

The obtained experimental data suggest a gradual change of the chemical bonding in the $Cu_xTiSe_2$ compounds with Cu concentration. Increase of the copper atoms concentration leads to compression of the Ti-Se sublattice and expansion of the Cu-Se sublattice in *c* direction and strengthening of the covalent component of the chemical bond. The concentration $x_0 = 0.33$ is the boundary between domination of the van der Waals (below $x_0$) and of the covalent (above $x_0$) bonding beetwen Cu and $TiSe_2$ lattice. Intercalation of copper atoms has a very weak effect on filling of the vacant Ti 3d states in the conduction band. Electronic structure of $Cu_xTiSe_2$ is mainly formed by the original TiSe$_2$ and the presence of free carriers is caused by copper atoms. It is shown that for the states of copper in Cu$_x$TiSe$_2$ the closest analogue in the ionic classification is Cu$^0$; at high copper concentrations the contribution from the Cu$^1$ states appears. Copper atoms form a diluted monatomic layer which is the source of the electrons in the conduction band and leads to the increase of the lattice parameters in the low copper concentrations region ($x < 0.33$). Measurement of Cu 3p-3d resonant spectra of the valence bands shows that Cu 3d band is almost full. The experimentally observed peaks below the Fermi level in the Ti 2p-3d and Cu 3p-3d resonant excitation modes indicate weak hybridization of copper and titanium 3d states and screening of titanium atoms by the electrons which take part in the dynamic charge transfer.

**ACKNOWLEDGMENTS**

The authors are grateful to ELETTRA synchrotron for support in the framework of CiPo and BACH beamlines. This work was performed within the framework of the Russian–German Laboratory at BESSY bilateral Program, project N 14201529. Work is supported by RFBR grant 14-03-00274 and by the project of the Multipurpose program of the Ural Division of RAS № 15-9-2-30.

Table. 1

Copper concentration dependence of the crystallographic parameters

*a, c* - lattice parameters; $z_{Se}$ – z-coordinate of the Se atom; d - interatomic distances; h – the height of the layer.

| x | a | c | $z_{Se}$ | $d_{(Ti-Se)}$ | $d_{(Cu-Se)}$ | $h_{(Se-Ti-Se)}$ | $h_{(Se-Cu-Se)}$ |
|---|---|---|---|---|---|---|---|
| 0.05 | 3.5408 | 6.0246 | 0.25773 | 2.56711 | 2.51186 | 3.1054 | 2.9192 |
| 0.1 | 3.5450 | 6.0423 | 0.2586 | 2.57498 | 2.51328 | 3.1251 | 2,9172 |
| 0.33 | 3.5631 | 6.0894 | 0.25672 | 2.58374 | 2.53506 | 3,1265 | 2,9629 |
| 0.6 | 3.5819 | 6.1126 | 0.24472 | 2.55231 | 2.59067 | 2.9918 | 3,1208 |
| 0.7 | 3.5807 | 6.1068 | 0.24431 | 2.54946 | 2.59074 | 2,9839 | 3,1229 |

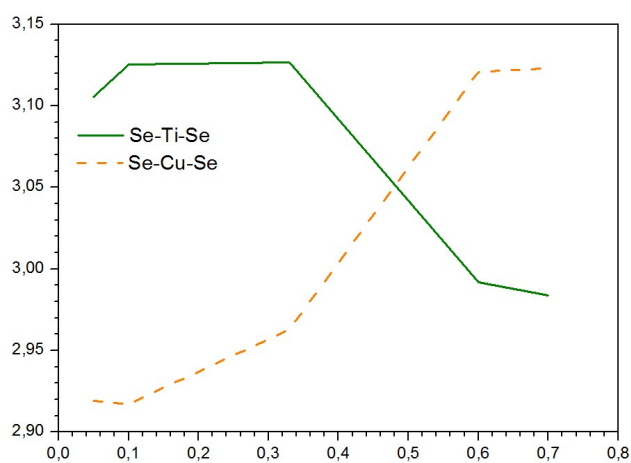

Figure 1. Width of Se-Ti-Se "sandwich" and Se-Cu-Se interlayer space width.

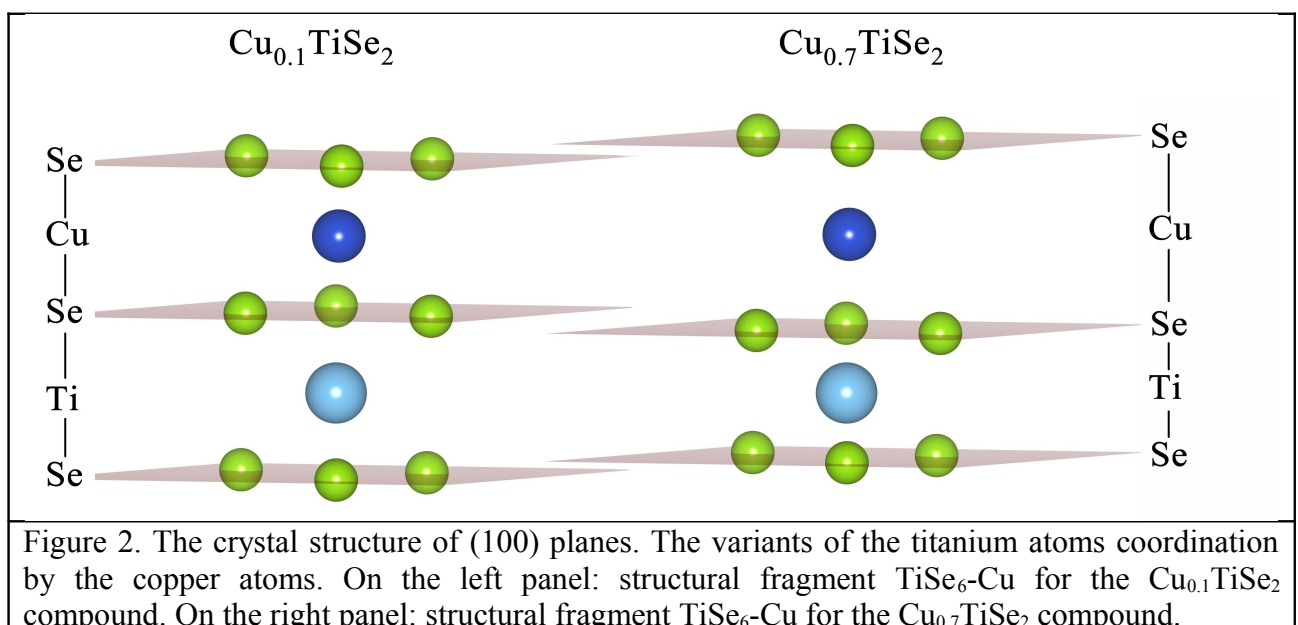

Figure 2. The crystal structure of (100) planes. The variants of the titanium atoms coordination by the copper atoms. On the left panel: structural fragment TiSe$_6$-Cu for the Cu$_{0.1}$TiSe$_2$ compound. On the right panel: structural fragment TiSe$_6$-Cu for the Cu$_{0.7}$TiSe$_2$ compound.

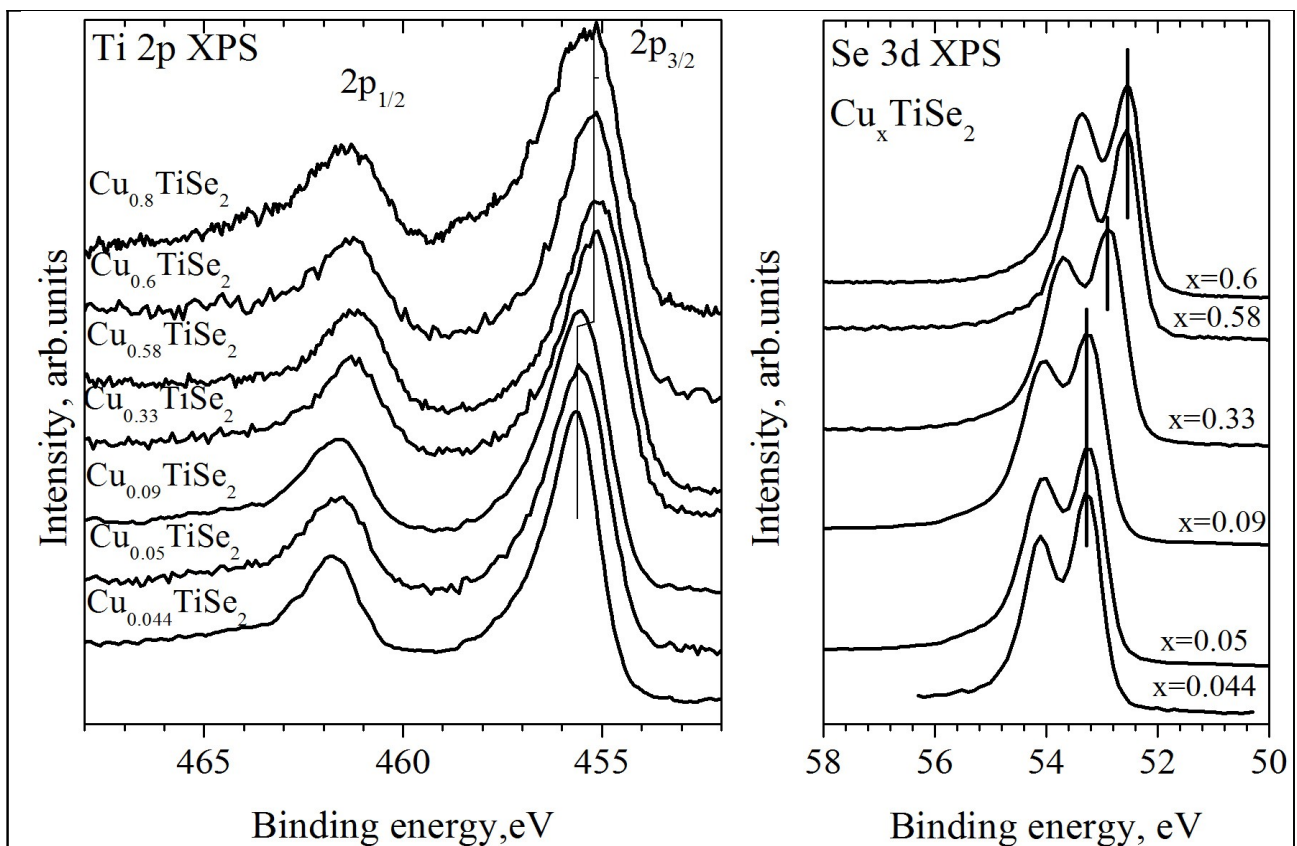

Figure 3. Left panel: core Ti 2p level spectra. Right panel: core Se 3d level spectra. The dashed line shows the spectrum of $Cu_{0.07}TiSe_2$ contaminated with $Cu_2Se$.

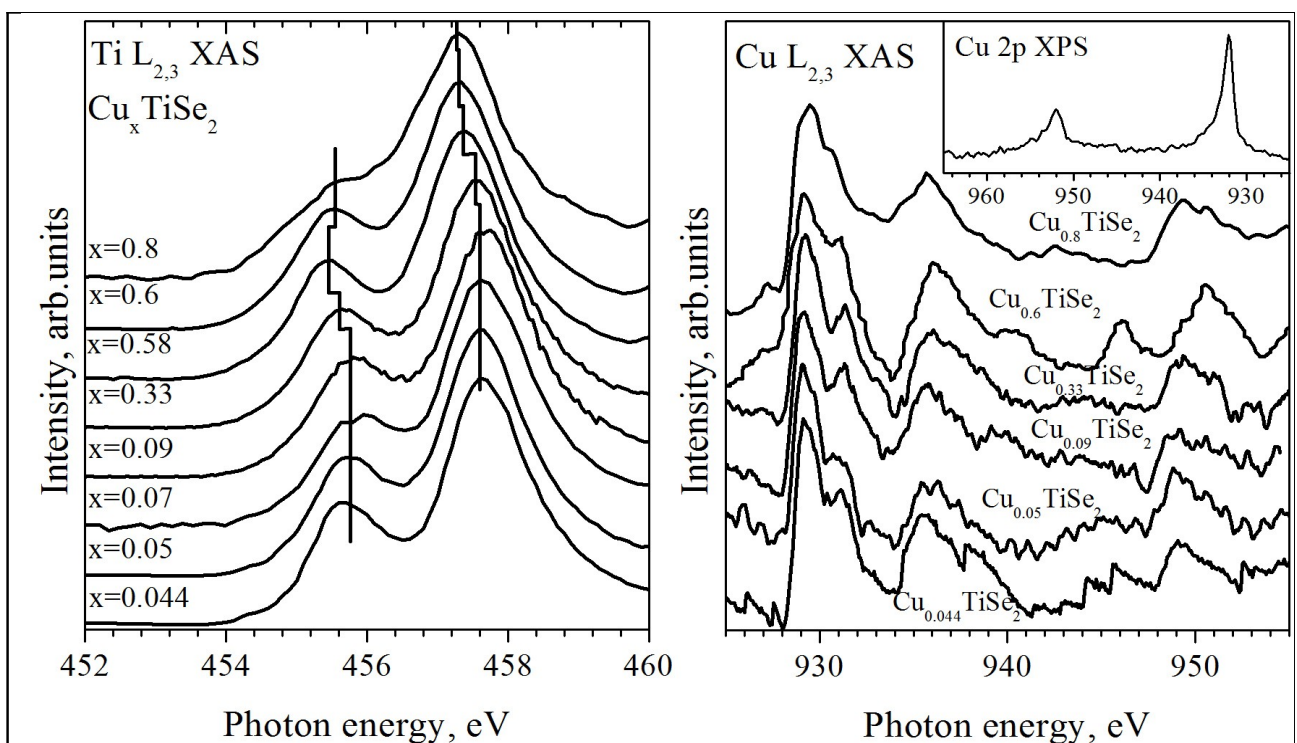

Figure 4. Left panel: Ti $L_3$ XAS of $Cu_xTiSe_2$. Right panel: Cu $L_{2,3}$ XAS of $Cu_xTiSe_2$. On the inset of right panel: typical core Cu 2p level spectrum in $Cu_xTiSe_2$.

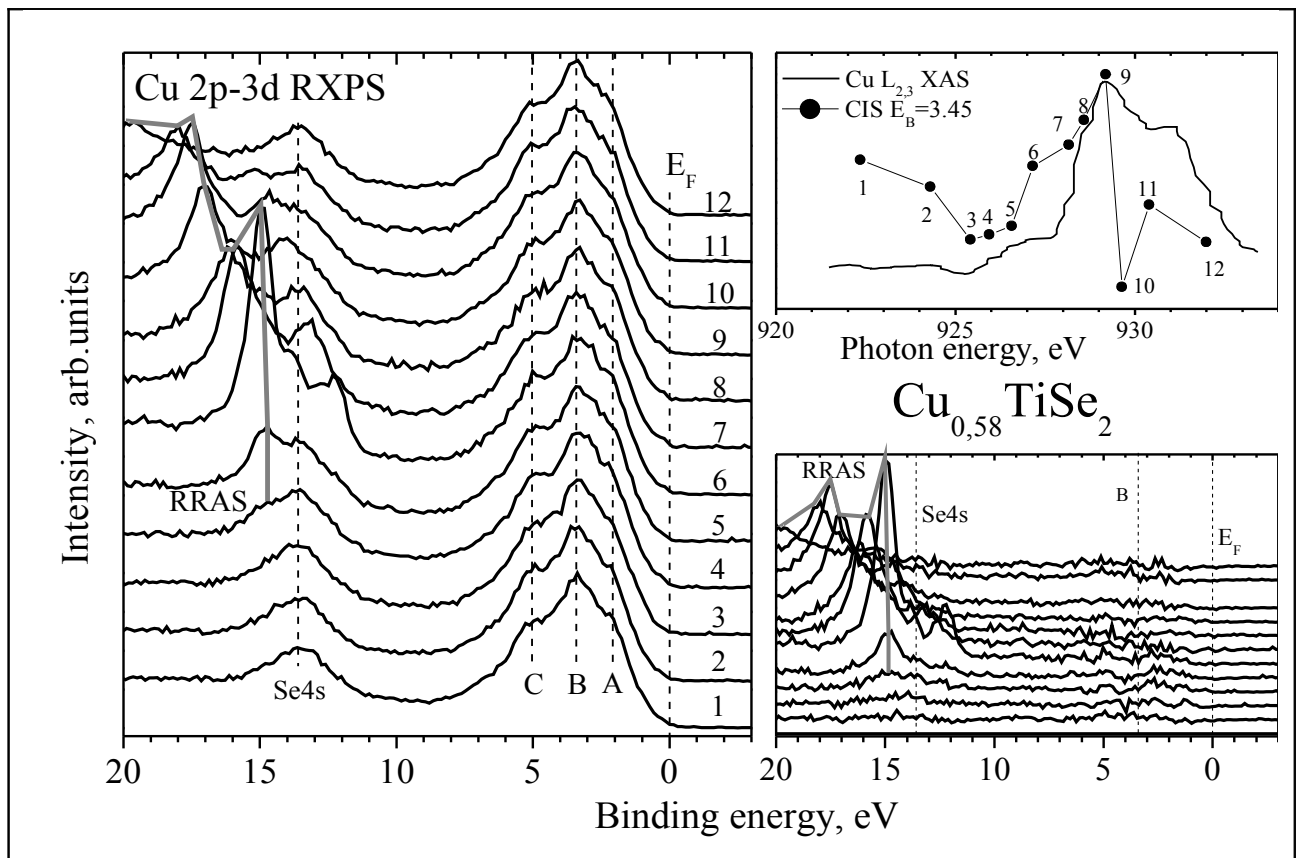

Figure 5. The valence band spectra in the Cu 2p-3d excitation mode for $Cu_{0.58}TiSe_2$. On the upper right charts the corresponding absorption spectra and CIS obtained at 3.45 eV binding energies from the valence band spectra are shown. On the bottom right charts the differential spectra obtained by substraction of the corresponding spectrum with minimal excitation energy from the spectra with other excitation energies.

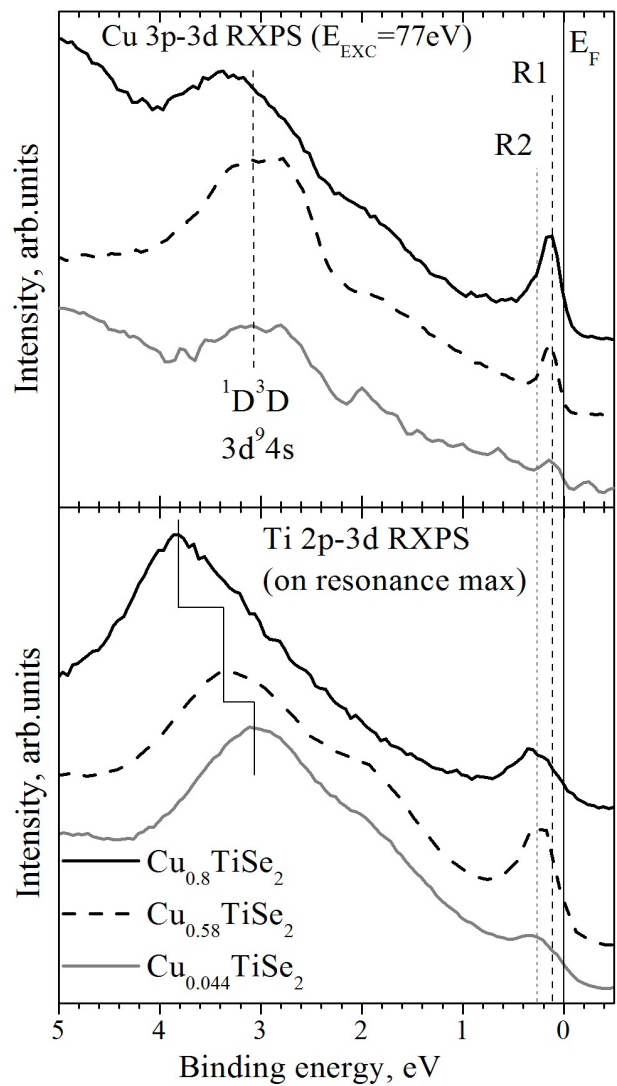

Figure 6. The valence band spectra of $Cu_{0.044}TiSe_2$, $Cu_{0.58}TiSe_2$ and $Cu_{0.8}TiSe_2$ near the Cu 3p-3d resonant excitation energy (upper panel) and the valence band spectra on the Ti 2p-3d resonance for the same compounds (bottom panel). For R1 $E_b = 0.11$ eV, for R2 $E_b = 0.27$ eV.